\documentclass[10pt]{iopart}
\usepackage{graphicx}
\expandafter\let\csname equation*\endcsname\relax
\expandafter\let\csname endequation*\endcsname\relax
\usepackage{amsmath}
\usepackage{amssymb}
\usepackage{caption}
\usepackage{subcaption}
\usepackage{listings}
\usepackage{color}
\usepackage{url}
\usepackage{mhchem}
\definecolor{codegreen}{rgb}{0,0.6,0}
\definecolor{codegray}{rgb}{0.5,0.5,0.5}
\definecolor{codepurple}{rgb}{0.58,0,0.82}
\definecolor{backcolour}{rgb}{0.95,0.95,0.92}

\lstdefinestyle{mystyle}{
    backgroundcolor=\color{backcolour},
    commentstyle=\color{codegreen},
    keywordstyle=\color{magenta},
    numberstyle=\tiny\color{codegray},
    stringstyle=\color{codepurple},
    basicstyle=\ttfamily\footnotesize,
    fontadjust=true,
    basewidth=0.5em,
    columns=fixed,
    breakatwhitespace=false,
    breaklines=true,
    captionpos=b,
    keepspaces=true,
    numbersep=5pt,
    showspaces=false,
    showstringspaces=false,
    showtabs=false,
    tabsize=2
}

\newcommand{\sn}[2]{\ensuremath{{#1}{\cdot}10^{#2}}}

\newcommand{\atkforcefield}{\mbox{\em ATK-ForceField}}
\newcommand{\atk}{{\em ATK}}
\newcommand{\atomistixtoolkit}{{\em Atomistix ToolKit}}
\newcommand{\vnl}{{\em VNL}}
\newcommand{\virtualnanolab}{{\em Virtual NanoLab}}
\newcommand{\tremolox}{{\em TremoloX}}
\newcommand{\atkpython}{ATKPython}

\begin{document}

\title[ATK-ForceField: A New Generation Molecular Dynamics Software Package]{ATK-ForceField: A New Generation Molecular Dynamics Software Package}

\author{Julian Schneider$^1$, Jan Hamaekers$^2$, Samuel T. Chill$^1$, S{\o}ren Smidstrup$^1$, Johannes Bulin$^2$, Ralph Thesen$^2$, Anders Blom$^1$, Kurt Stokbro$^1$}

\address{$^1$ QuantumWise A/S, Fruebjergvej 3, DK-2100 Copenhagen, Denmark}
\address{$^2$ Fraunhofer Institute for Algorithms and Scientific Computing SCAI, Schloss Birlinghoven, 53754 Sankt Augustin, Germany}
\ead{kurt.stokbro@quantumwise.com}
\vspace{10pt}

\begin{abstract}
    \atkforcefield{} is a software package for atomistic simulations using
    classical interatomic potentials. It is implemented as a part of the
    \atomistixtoolkit{} (\atk{}), which is a Python programming environment that
    makes it easy to create and analyze both standard and highly customized
    simulations. This paper will focus on the atomic interaction potentials,
    molecular dynamics, and geometry optimization features of the software,
    however, many more advanced modeling features are available. The
    implementation details of these algorithms and their computational
    performance will be shown. We present three illustrative examples of the
    types of calculations that are possible with \atkforcefield{}:
        modeling thermal transport properties in a silicon germanium crystal,
        vapor deposition of selenium molecules on a selenium surface,
        and a simulation of creep in a copper polycrystal.
\end{abstract}


\section{Introduction}

Molecular dynamics (MD) simulations have become a versatile and widely used
tool in scientific computer simulations \cite{allen1989computer, Griebel2007}.
MD can be used to gain insight in atomic-scale processes when experimental
techniques are unable to provide sufficient resolution. MD is used to sample
microscopic ensembles, to study microscopic dynamical and transport
properties, such as diffusion or thermal transport, or to simulate physical
processes in order to access atomic structures for which little
experimental information is available, e.g. structures of complex interfaces,
amorphous materials, or vapor-deposited films.

MD simulations are often used to interpret and understand the results of
experimental studies, but they are also capable of predicting or screening new
materials or processes which can significantly reduce the number and cost of
experiments \cite{Hong2015, Okimoto2009}.

Numerical optimization algorithms are another, closely related, core tool in
atomic scale modeling, as locating stable structures and reaction mechanisms
can be formulated as optimization problems. Minimum energy structures are
obtained by performing a local minimization of the potential energy (or
enthalpy for solids under isotropic pressure). Reaction mechanisms can be
located by minimizing the projected forces of a nudged elastic band (NEB)
\cite{Henkelman2000}.

A wide variety of software packages are available for MD and optimization.
Some packages like LAMMPS \cite{PLIMPTON1995}, DL\_POLY \cite{DLPOLY}, or MBN
Explorer \cite{MBNExplorer}, are designed to work efficiently with a wide
spectrum of materials, however, most are optimized for certain classes of
systems, such as crystals (e.g. VASP \cite{VASPKresse} and GULP
\cite{GULPGale2003}), biomolecules (e.g. AMBER \cite{AMBERCase}, GROMACS
\cite{GROMACSAbraham2015}, CHARMM \cite{NAMDPhillips}, and NAMD \cite{NAMD}),
or polymers (e.g. ESPResSo \cite{ESPRESSOLimbach}). Almost all of these
packages only support either empirical potentials or ab-initio methods. For many
tasks it may be desirable to quickly and seamlessly switch between
different levels of accuracy, e.g. when validating or parameterizing empirical
potentials against higher-level calculations, or when using classical
potentials to obtain mechanical and thermal properties and DFT for electronic
properties of the same system \cite{Markussen2009, Markussen2017}.
Yet, very few (e.g. CP2K \cite{CP2K} or ADF \cite{ADF}) simulation codes
provide forces and energies from both of ends of the atomic accuracy spectrum.
The atomic simulation environment (ASE) \cite{ASE} addresses this problem by
integrating a broad range of atomic-scale simulation methods into a common
Python framework which implements various tools for atomic scale simulations,
including molecular dynamics and optimization functionality. However, the ASE
is not optimized towards highly efficient classical MD and is therefore mainly
used in combination with DFT methods.

Some physical problems require very specific techniques which reach beyond the
standard algorithms commonly implemented in most MD codes or may even need to
be custom-tailored to the application at hand.
In this case, a flexible way of extending and combining the standard simulation
algorithms is desired. Ideally this task should be done as seamlessly as
possible, without having to manually interface different programs, programming
languages, or converting input and output files from one format to the another.
This concerns not only the simulation technique itself, but, equally important,
the subsequent analysis and visualization of the simulation trajectory and
other raw data. Although most major software packages come with a more or less
extensive suite of analysis and visualization tools, these are often included
as small standalone programs, which have to be compiled and invoked
individually. This often leads to each user developing a custom workflow by
patching together different tools, which significantly impedes reusability and
reproducability.

As MD becomes a more important and commonly used technique, it is important to
make it accessible to a broader spectrum of scientists. Thus there is a need
for user-friendly ways of setting up and analyzing simulations. This becomes
particularly important in industrial settings, where the time a researcher
needs to become an efficient user of a new code impacts the business'
bottom line.

In this paper, we describe how \virtualnanolab{} (\vnl{}) and the
\atomistixtoolkit{} (\atk{}) developed by QuantumWise are designed to meet these
challenges. Generally, the main goal of VNL and ATK is to make atomistic
simulation techniques easily accessible to an increasing number of researchers.
To this aim, the VNL provides a graphical user interface, in which all steps
involved in the workflow of atomistic simulations, i.e designing the
configuration to be simulated, setting up and running the actual calculation,
as well as analyzing and visualizing the results, can be carried out.

ATK provides a Python scripting interface to all the calculators and
simulation methods. {\em ATKPython} is inspired by ASE \cite{ASE}, however, it is more
extended and rich in functionality and specifically with respect to MD
simulations there is a negligible overhead from the Python layer. It is designed
in a modular way, meaning that individual components can easily be exchanged
without having to change the entire workflow. For example, switching from an
empirical potential to DFT, only requires a single line change in the input
script. Most importantly, by using Python, it is possible to for the user to
implement custom simulation techniques without having to re-compile the code.

For the scope of this paper we will specifically focus on the \atkforcefield{}
package and the dynamics and optimization module in ATK, although, in
principle all scripts and examples could as well be done with DFT methods.
ATK-ForceField is the calculation engine for empirical potentials. It mainly
consists of the \tremolox{}-calculator, which is developed by the Fraunhofer
Institute for Algorithms and Scientific Calculations (SCAI). The TremoloX-
calculator is based on SCAI's {\em Tremolo-X} software package \cite{TremoloX}
for numerical simulation in molecular dynamics and mechanics
\cite{Griebel2007}. The TremoloX-calculator is optimized for
large scale simulations, provides state-of-the-art scalable fast methods for
long-range interactions \cite{PhysRevE.88.063308} and has successfully
performed in several applications, ranging from nanoparticles and
nanocomposites over cementitious materials to electrolytes and
biomolecules~\cite{Frankland.Caglar.Brenner.ea:2002,Griebel.Jager.Voigt:2004,Griebel.Hamaekers:2004,Griebel.Hamaekers:2005,Dolado.Griebel.Hamaekers:2007,Bittner.Heber.Hamaekers:2009,Griebel.Hamaekers.Heber:2007,Dolado.Griebel.Hamaekers.ea:2011,Neuen.Griebel.Hamaekers:2012,Diedrich.Dijkstra.Hamaekers.ea:2015}.
TremoloX itself is implemented in parallel, although the ATK-ForceField currently
only supports the shared-memory OpenMP version.
The combination of the ATK-ForceField calculator module and the dynamics module in ATK
provides state-of-the-art MD functionality and optimization algorithms, as
well as more advanced techniques such as nudged elastic band (NEB) and
adaptive kinetic Monte Carlo (AKMC) simulations.

Although in principle VNL and ATK are commercial software packages, VNL
and ATK-ForceField are freely available for academic groups. We hope that in
this way we can develop an academic community of end users which will develop
new application examples for the \atkpython{} framework and new plugins for
VNL. In fact, all examples presented in section \ref{sec:examples} of this
paper can be run under this free license.

In the first few sections, we describe the basic concepts of modeling atomic
geometries and how interatomic potentials are composed and calculated inside
ATK-ForceField, as well as some performance benchmarks. The next section reviews
the basic concepts of MD and optimization, and their implementation in ATK to
meet the specifications defined above. Subsequently, the simulation and
analysis workflow in VNL and ATK is presented. We then describe the
integration of the calculation engine into the Python platform, including some
small code samples. In the following section, we present some example simulations to
demonstrate the capability of MD using ATK. Finally, we give an overview for
the future development of ATK-ForceField.

\section{Atomic Configurations}
\label{sec:geometry}

\begin{figure}
\begin{center}
    \centering
    \includegraphics[width=0.9\textwidth]{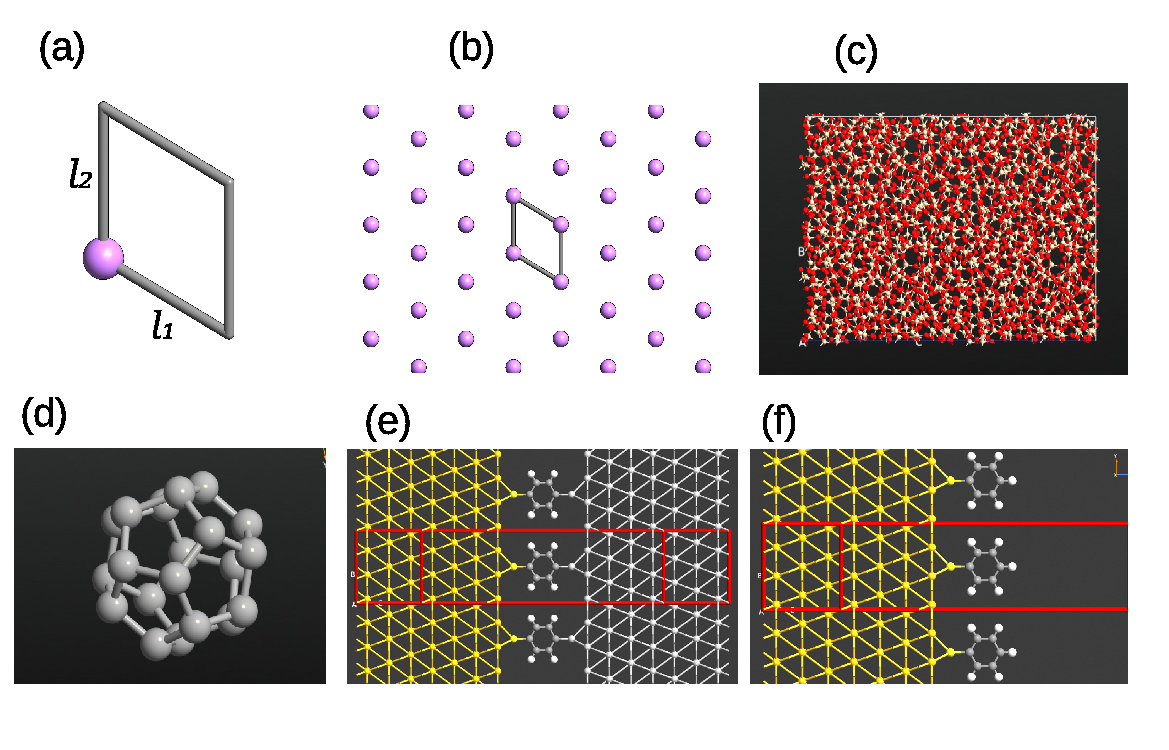}
    \captionof{figure}{Examples of configurations in VNL-ATK:
    {\it (a)} Bulk configuration of the unit cell of a 2D lithium crystal, and {\it (b)} its periodic extension.
    {\it (c)} Bulk configuration of a supercell containing an amorphous silica.
    {\it (d)} Molecule configuration of a fullerene molecule.
    {\it (e)} Device configuration consisting of left electrode, central region and right electrode.
    The transport direction is from left to right and the structure is periodic in the two directions perpendicular to the transport direction.
    {\it (f)} Surface configuration, being a device configuration with only the left electrode.
    An electric field can be applied to the surface through the right boundary condition.
    }
    \label{fig:configurations}
\end{center}
\end{figure}

At the beginning of each calculation the system under investigation
must be specified. ATK offers several configuration types for different applications.
The simplest case is a \emph{molecule configuration} which describes a system containing an isolated
collection of atoms with their positions $\mathbf{r}_i$ and elements $Z_i$
(see figure \ref{fig:configurations} (d)).
For systems like crystals which are are inherently periodic and require a
different treatment, ATK offers the so-called \emph{bulk configuration} class.
A bulk configuration contains the central cell and repeats it periodically in
all directions during simulations (see figure \ref{fig:configurations} (b)).
This simulation cell can be a unit cell, e.g. of a crystal (figure
\ref{fig:configurations} (b)), but it can also be used as a supercell,
containing systems without a pronounced periodic character, such as e.g.
amorphous, liquid, or interface systems, as well as slab geometries (figure
\ref{fig:configurations} (c)). In ATK, the three lattice vectors
$\mathbf{l}_i$ spanning the cell can either be given explicitly, as a cell
matrix $\mathbf{L} = [\mathbf{l}_1, \mathbf{l}_2, \mathbf{l}_3]$, or by using
one of the predefined classes that represent the 14 different Bravais lattice
systems, which require only the independent lattice parameters to be
specified.

Apart from bulk and molecule configurations, ATK also offers so-called \emph{device
(two-probe) configurations}, which consist of a central region, as well as a
left and right electrode. A device configuration is periodic in two
directions, whereas the third dimension is bounded by two infinite electrodes
(see Fig. \ref{fig:configurations} (e)). The central region and
electrode parts are each defined as bulk configurations, and, to be a valid
definition, the left and right part of the central region must be identical to
the corresponding adjacent electrode. These configurations can be used to
simulate electron or phonon transport via the non-equilibrium Green's functions
(NEGF) technique \cite{Brandbyge2002}. Finally, to accurately study surface
chemistry, ATK provides \emph{surface (one-probe) configurations}, which are device
configurations with only one electrode (see Fig. \ref{fig:configurations} (f)),
accurately describing a semi-infinite surface geometry beyond the common
slab supercell approximation.

\section{Interatomic Potentials}
\label{sec:potentials}

In general, the underlying model to describe the dynamics of an atomic
systems is the full Schr\"odinger equation for the electrons and
nuclei under consideration. Its high-dimensionality, however, makes a
direct numerical treatment impossible and thus one has to resort to
model approximations.
A hierarchy of these approximations is typically necessary to calculate the
properties of nano-scale systems. Based on the Born-Oppenheimer formulation,
which disentangles the motions of the nuclei and the electrons into a classical and a
quantum mechanical part \cite{Tadmor2011,Griebel2007}, the solution to the electronic Schr\"odinger equation
is commonly approximated by density functional theory (DFT) or tight-binding (TB) techniques,
and a variety of such methods is readily available in the ATK program package.

Although these QM methods have become very powerful, the accessible systems
sizes are still not sufficient for a realistic simulation of many nano-structured
systems of practical interest.
To this aim, the idea is to determine an approximation $V$ to
the exact potential energy surface in an efficient analytical
functional form, a so-called force field or {\em classical potential}.


To overcome the curse of dimensionality of $V$ being a function of all atomic positions,
classical potentials are typically based on an appropriate many-body expansion, i.e.\
\begin{equation*}
  V(\vec{x}) = V_\emptyset +
  \sum_{i}V_{\{i\}}(\mathbf{r}_i)+\sum_{i<j}V_{\{i,j\}}(\mathbf{r}_i, \mathbf{r}_j) + \ldots +
  V_{\{1,\ldots,N\}}(\mathbf{r}_1,\dots,\mathbf{r}_N).
\end{equation*}
If the size of the terms decays fast with e.g. the order $k=|u|$ of the $k$-body contributions $V_u$,
then a proper truncation of the expansion results in a substantial
reduction in computational complexity.

A very simple class of classical potentials are two-body potentials
which have the form
\begin{equation}
   {V}_2(\vec{x}) = \sum_{i=1}^N \sum_{j\neq i} v_{i,j}(\mathbf{r}_i, \mathbf{r}_j), \label{twobody}
\end{equation}
where $v_{i,j}$ is a function that depends only on the properties of
two particles. A variety of well-known potential classes, such as the Lennard-Jones
potential~\cite{Jones1924} or
the Buckingham potential~\cite{Buckingham1938} can be written in the form of
equation~(\ref{twobody}). In the periodic case,
interactions between the particles in the unit cell and all other
particles, including their periodic copies, are taken into account:
\begin{equation}\label{twobodyperiodic}
   {V}_2(\vec{x}, L) = \sum_{i=1}^N \sum_{\mathbf{n}\in\mathbb{Z}^3} \sum_{j=1\atop \mathbf{r}_i\neq L\mathbf{n} + \mathbf{r}_j}^N  v_{i,j}(\mathbf{r}_i, L\mathbf{n} + \mathbf{r}_j).
\end{equation}
In a similar fashion, three-body potentials
and straightforwardly higher-order $k$-body potentials can be defined, where the
periodic case can be treated analogously to two-body potentials
\cite{Griebel2007,Griebel.Hamaekers:2005b}. An example for a widely
used three-body potential -- in combination with two-body terms -- is
the Stillinger-Weber potential~\cite{Stillinger1985}.

If the interaction terms $v_{u_1, \ldots, u_k}$ are of short range, meaning that the error that results
from discarding terms involving particles further apart than a cutoff
distance $r_{cut}$ can be made arbitrarily small by choosing $r_{cut}$
sufficiently large
\cite{frenkel1996understanding}, the total potential function at any fixed $\vec{x}$, involves only $O(N)$
terms.
%
Most interaction potentials applied within an approximation potential
of the electronic energy $V_{el}$ are of short range.
However, in particular Coulomb interactions, i.e.
\begin{equation}
V_{ij}(r) = \frac{q_i q_j}{r}
\end{equation}
or dipolar interactions do not satisfy the short range condition.
Hence, the simple cutoff approach can not be applied and the interactions with
all periodic images need to be taken into
account~\cite{frenkel1996understanding}. To this end, a long range
potential is typically decomposed in a short range but non-smooth part
and a long range but smooth part which leads to methods with costs of
order $O(N^{\frac{3}{2}})$ and even of order $O(N\log(N)^\alpha)$,
where $\alpha \leq 0$ depends on the specific
method~\cite{Griebel2007,toukmaji1996ewald,Essmann1995,Fennell2006,PhysRevE.88.063308}.

Besides a variety of standard two- and three-body potentials,
ATK-ForceField implements a number of bond-order potentials or cluster
functionals~\cite{Tadmor2011}, like Abel and
Tersoff~\cite{Abell1985,Tersoff1988}, embedded atom method (EAM) and modified EAM (MEAM) potential
types~\cite{DawBaskes1984,baskes1992modified} or the Sutton-Chen
potential~\cite{Sutton1990}.  Here, the expansion contributions
$V_{u}$ do not depend on the particles $\mathbf{r}_{u_1}, \ldots,
\mathbf{r}_{u_k}$ alone, but in addition also on their neighboring
environment.

A further class of more advanced potential models implemented in ATK-ForceField,
are potentials in which the partial charges of the particles are not fixed but
are calculated by minimizing the potential energy with respect to these charges:
\begin{equation}\label{equ:nonlinQEQ}
   V(\mathbf{r}_1,\dots,\mathbf{r}_N) = \min_{q_1,\dots, q_N} \tilde V(\mathbf{r}_1,\dots,\mathbf{r}_N, q_1,\dots, q_N).
\end{equation}
This includes all charge-optimized many-body (COMB) potentials~\cite{Yu2007} and ReaxFF
potentials~\cite{Vanduin2001,chenoweth2008reaxff}, as well as ReaxFF+,
a variant of ReaxFF, which combines charge equilibration with the
bond-order principle to better distinguish between covalent and ionic bonds~\cite{Boehm2016}.
The equilibrium charges are obtained by using a nonlinear optimization algorithm.
A similar ansatz is chosen for
potentials with induced dipoles where the dipoles are once again
calculated by minimizing the potential energy with respect to these
dipoles. Induced dipoles are used in the third-generation COMB potentials
\cite{Liang_COMB3} and in the aspherical ion model potential~\cite{Aguado2004}.

Moreover, to describe ionic polarizability in ionic materials
ATK-ForceField includes also the widely used core-shell potential
models~\cite{mitchell1993shell}. Here, in contrast to the minimization
problem (\ref{equ:nonlinQEQ}), the potential energy is minimized with
respect to the position of charged shells.

For a wide range of applications, it is sufficient to consider molecular
systems which retain a static bond topology during simulation. A common
example for these kind of force fields are biomolecular forces fields, which,
apart from basic Lennard-Jones and Coulomb interactions, typically involve
potentials to model bond stretching, bond angles, dihedral torsions, and in
some cases improper torsions (see e.g. \cite{Cornell1995,
mackerell2004empirical}).  In contrast to the non-bonded potentials,  these
bonded potentials only act between atoms which are connected via the bond
topology. The same holds for valence force fields (VFF), which use similar
bonded potential terms and which are typically used to describe elastic and
vibrational properties in covalent solids~\cite{keating1966effect}.

A summary of the main included potential model types is given in
Table~\ref{tab:potentials}. Several hundred pre-defined parameter
sets for these potentials are currently provided in ATK-ForceField
\footnote{For a list of most provided parameter sets e.g.\\
\url{http://docs.quantumwise.com/manuals/atkforcefield.html\#tremolox-potential-parameter-sets} \\
or \url{http//www.tremolo-x.com/en/tremoloxpotentials.html}}.
\begin{table*}
  \caption{Selected potential models included in ATK-ForceField.}
  \label{tab:potentials}
  \centering
  \begin{tabular}{l|p{6cm}|l}
    Potential model & Special properties & Reference\\\hline
    Stillinger-Weber & three-body & \cite{Stillinger1985}\\
    Embedded atom model (EAM) & many-body & \cite{mishin2001structural}\\
    Modified embedded atom model (MEAM) & many-body\newline directional bonding & \cite{Baskes1997}\\
    Tersoff & bond-order & \cite{Tersoff1988}\\
    ReaxFF  & bond-order\newline dynamical charges & \cite{chenoweth2008reaxff}\\
    COMB/COMB3 & bond-order\newline dynamical charges\newline induced dipoles & \cite{Yu2007}\\
    Core-shell & dynamical charge fluctuations & \cite{mitchell1993shell}\\
    Tangney-Scandolo  & induced dipoles & \cite{tangney2002}\\
    Aspherical ion model & induced dipoles and quadrupoles\newline dynamical ion distortion & \cite{Rowley1998}\\
    biomolecular and valence force fields & static bonds  & \cite{mackerell2004empirical, keating1966effect}\\
  \end{tabular}
\end{table*}
In some applications combinations of different
potential models are required to provide an appropriate description of the considered
atomic systems.
Apart from a few exceptions, ATK-ForceField allows arbitrary combinations of
different potentials.
Moreover, if a specialized pair potential form is not implemented in ATK-ForceField,
one can always resort to a tabulated representation of the potential.

Since ATK originates from an ab-initio based code, a major field of application is
the simulation of small crystal unit cells. Thus, special emphasis has been
placed on implementing ATK-ForceField such that it correctly calculates the
interactions in cells much smaller than the interaction range of the
potential, by including as many periodic images as necessary
\cite{Griebel.Hamaekers:2005b}, but at the same time efficiently handles
simulations of large systems without loss of performance.

\section{Performance}
\label{sec:performance}

\begin{figure}
\begin{center}
  \includegraphics[width=0.7\textwidth,clip]{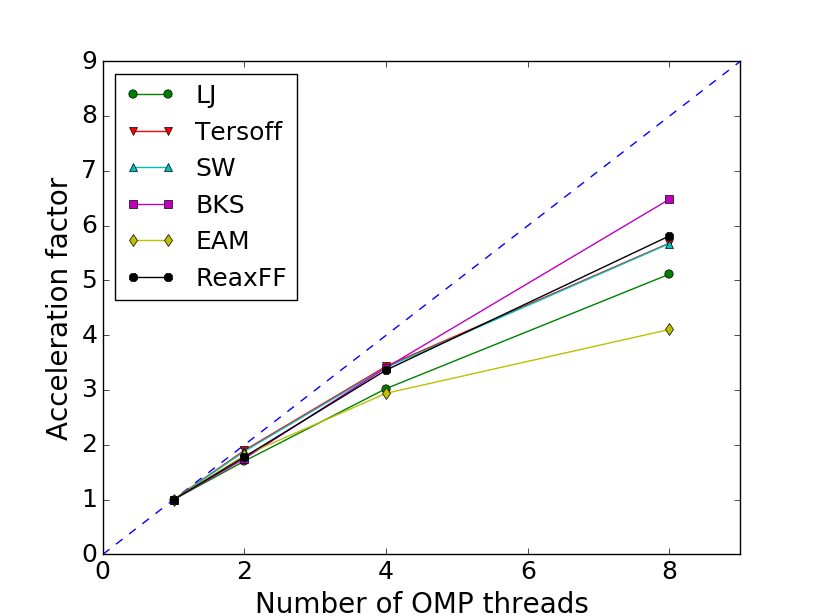}
  \caption{Acceleration factor of various potentials when parallelized
    over different number of cores of a single compute node using
    OpenMP threading.  The ideal scaling is depicted by the dashed
    line. All system contain around 32 000 atoms.}
  \label{fig:performance}
\end{center}
\end{figure}

In this section we will present the performance of MD simulations with
different potentials in ATK-ForceField. Let us first review
some implementation details of the ATK-ForceField back-engine
Tremolo-X. Similar to other performance-intensive parts in ATK,
the Tremolo-X back-engine is implemented in C and C++.
All short range potential terms are
computed with linear cost complexity $O(N)$ based on the combination
of the well-known Verlet list and cell linked list
algorithms~\cite{hockney1988computer,allen1989computer}. In this
hybrid approach, the cell linked list technique is used to setup and
update (when necessary) the Verlet pair lists, which are then used to
efficiently compute all short range potentials. To improve the memory
hierarchy performance a Hilbert-Peano curve data ordering can be
applied~\cite{mellor2001improving,Griebel2007}. To speed up the
evaluation all short range two-body potential terms can be
automatically aggregated and tabulated for all pairs of particle
types, respectively. To compute long-range Coulomb interactions in the
periodic case (i.e.\ {\em BulkConfiguration}), we implemented e.g. the
conventional Ewald summation~\cite{Ewald1921}, the smooth particle
mesh Ewald~\cite{Essmann1995} and the DSF approach~\cite{Fennell2006},
with cost complexities of order $O(N^\frac{3}{2})$, $O(N\log(N))$ and
$O(N)$, respectively. In the case of non-periodic boundary conditions
(i.e.\ {\em MolecularConfiguration}), we implemented a fast multipole
method (FMM)~\cite{greengard1989evaluation} with linear cost
complexity $O(N)$.

Tremolo-X itself is implemented in parallel using a hybrid distributed (MPI) and
shared memory (OpenMP) approach. Currently, ATK-ForceField only provides the
shared memory OpenMP version and in the following we limit our performance
study to computations on a single node using OpenMP. The simulations were run on
an Intel E5-2687W 3.1GHz CPU with 8 cores.

Six different potentials were benchmarked.
We start with a simple Lennard-Jones (LJ) system, built by
Argon atoms on a slightly randomized lattice with a particle density of 0.84,
given in units of the LJ-minimum distance $\sigma$.  Such a system is often
used as benchmark system for the performance of force-field MD
simulations \cite{PLIMPTON1995}.  We examine a bulk configuration of
a fcc-silicon crystal, which is simulated using a Stillinger-Weber (SW)
potential \cite{Stillinger1985} and a Tersoff potential \cite{Tersoff1988}.  As
an example for a system with electrostatic interactions we simulate a SiO$_2$
cristobalite crystal, using the potential proposed by van Beest, Kramer, and
van Santen (BKS) et al. \cite{vanbeest1990}. The electrostatic interactions are
calculated via the SPME algorithm using a cutoff of 9~\AA, whereas the
dispersion interactions are truncated at 10~\AA.  To benchmark the performance
of potentials based on the embedded-atom-method (EAM), we simulate a copper
crystal, using the potential of Mishin et al. \cite{mishin2001structural}.
Finally, we consider an even more complex, reactive ReaxFF-potential.  Here, we
simulated a triaminotrinitrobenzene (TATB) crystal, and we employed the
potential of Strachan et al. \cite{strachan2003shock}, which was designed for
such systems.  All system comprise around 32 000 atoms.  The simulations were
run for 100 MD steps and the total wallclock time $T$ for the MD loop is measured.
All simulations use a time step of 1 fs, except the ReaxFF simulation which used a
time step of 0.2 fs.

Figure \ref{fig:performance} shows the acceleration factor $T(1)/T(N)$
with respect to the simulation time on a single CPU. The ideal scaling
behavior is depicted by the dashed line.  Almost all potentials show
good scaling on up to 8 threads. Even the complicated ReaxFF-potential
scales very well in this range. Only the EAM potential exhibits a
scaling behaviour which deviates somewhat more from the ideal speedup.

The absolute times of these potentials are given in Tab. \ref{tab:timings},
extended by the some more elaborate potentials, such as the COMB \cite{Yu2007}
and Tangney-Scandolo (TS) \cite{tangney2002} potential. For both potentials
the SiO$_2$ cristobalite system was used with a time step of 0.2~fs. These
simulations were run on a single core of a laptop computer with an Intel
i7-3740QM 2.7 GHz processor. The time values are normalized to one atom and a
single MD step. Although a direct comparison absolute timings is always
difficult, these values are very similar to other state of the art simulation
codes such as LAMMPS~\cite{PLIMPTON1995}\footnote{\url{http://lammps.sandia.gov}}.

\begin{table*}
\caption{\label{tab:timings}
    Absolute timings for the MD runs using a single core with different potentials,
    normalized to one atom and a single MD step.}
    \begin{center}
        \begin{tabular}{l|cccccccc}
        \hline
        & LJ & Tersoff & SW & BKS & EAM & ReaxFF & COMB & TS \\\hline
        T (s)      & \sn{1.5}{-6} & \sn{7.7}{-7} & \sn{3.3}{-7} & \sn{2.8}{-5} & \sn{2.3}{-6} & \sn{1.9}{-4} & \sn{1.7}{-4} & \sn{3.6}{-4} \\
        \hline
        \end{tabular}
    \end{center}
\end{table*}

\section{Molecular Dynamics}
\label{sec:dynamics}

The ATK-ForceField package is designed and implemented to work highly efficiently
with ATK's molecular dynamics and optimization module. In this
section we briefly review the main concepts and implementation details behind molecular dynamics
in ATK, while in the next section we will discuss static approaches.

Essentially, molecular dynamics is a numerical integration to solve Newton's
equations of motion, which is typically performed by discretizing the time
into small time steps $\Delta t$.
The most common way to achieve this, is the velocity-Verlet algorithm \cite{Swope1982}.
All MD methods in ATK are based on this integration scheme.
The resulting trajectory $\{\mathbf{r}, \mathbf{v}\}(t=0, ..., T_{\rm final})$
represents the motion of the particles in the microcanonical (NVE) ensemble,
starting from the given initial conditions.
To simulate physical processes and experiments as precisely as possible, one
can additionally include external influences, such as coupling to a heat bath
or applying pressure or stress, which results in different ensembles.


ATK implements several algorithms for such thermostats and barostats. The preferred
types for weak coupling are the chained Nos\'{e}-Hoover thermostat
\cite{Martyna1992}, as well as the barostat proposed by Martyna et al.
\cite{Martyna94} for isotropic and anisotropic pressure coupling. Both
integrators have been implemented based on the algorithm proposed by Tuckerman
et al. \cite{Tuckerman06}. These methods typically yield stable trajectories
for a wide range of conditions and systems, and they exactly reproduce the
canonical ensemble. For completeness, ATK also provides a Berendsen thermostat
and barostat \cite{Berendsen84}, although these methods only approximately
reproduce the canonical ensemble.
For tight temperature coupling ATK offers an impulsive version of the Langevin
thermostat as described in Ref.~\cite{Goga12}. All thermostats in ATK can
optionally be used to heat and cool the system by specifying a linear increase
or decrease of the reservoir temperature over the simulation time. In contrast
to the force calculation, the integrators are implemented in Python, which can
be efficiently accomplished in a vectorized manner using the Numpy package
\cite{numpy_ascher}.

\begin{figure}
\begin{center}
  \includegraphics[width=0.7\textwidth,clip]{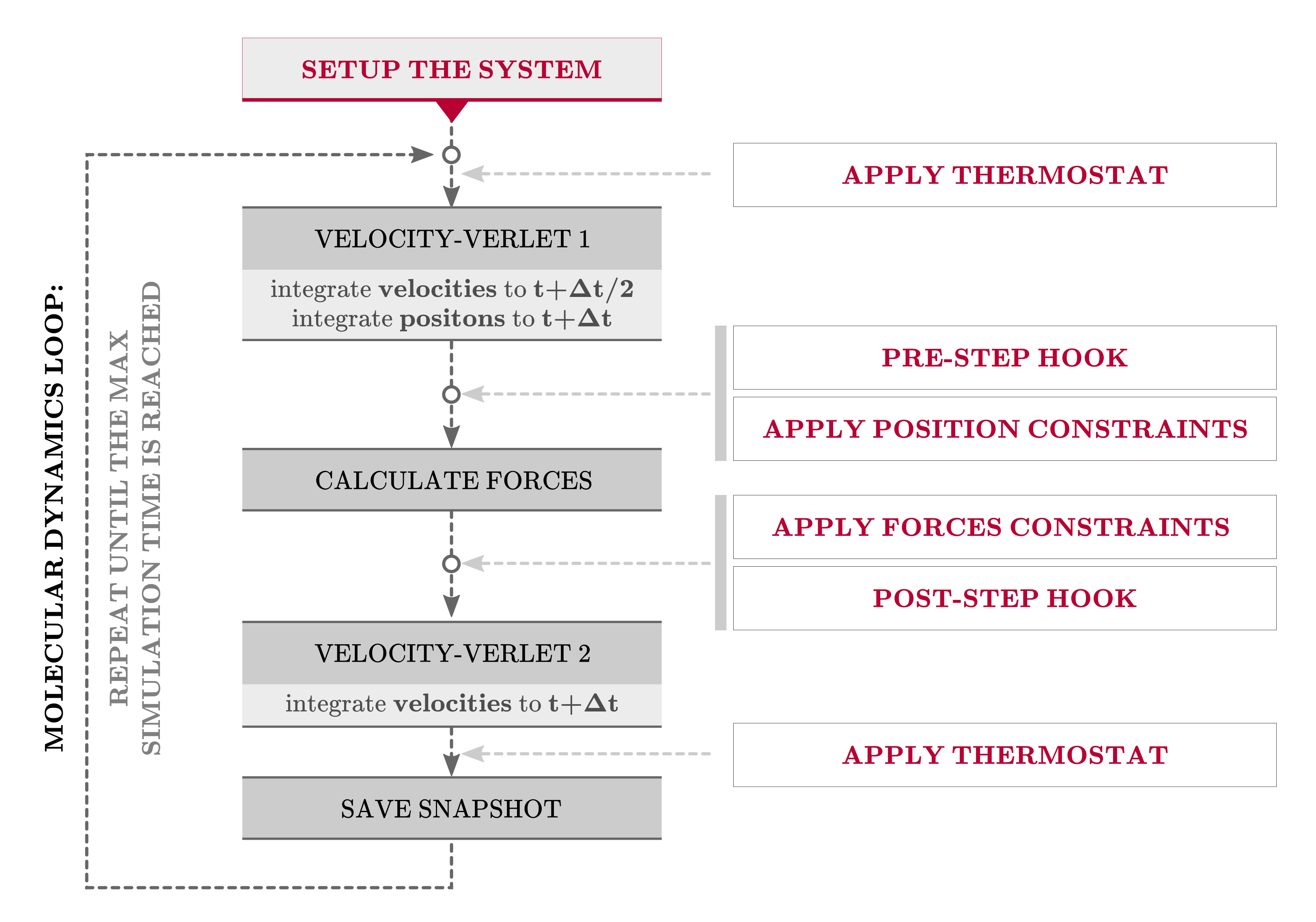}
  \caption{Flowchart of a typical MD loop in ATK.}
  \label{fig:mdloop}
\end{center}
\end{figure}

The essential functional blocks in a typical MD loop in ATK are schematically
depicted in Fig. \ref{fig:mdloop}.
One of the powerful features of ATK is that the MD loop works in the same way with
any calculator that is attached to the given configuration. That means,
instead of the ATK-ForceField calculator one could for instance employ a DFT or
a semi-empirical calculator, to achieve more accurate results at higher
computational cost, without having to change the MD block of the simulation.

To achieve more flexibility with respect to specialized simulation techniques,
the MD loop in ATK provides the possibility of defining {\it hook functions},
which interface with the simulation at specified points during each integration
loop (cf. Fig. \ref{fig:mdloop}). This way one can easily employ pre-defined or
user-defined custom operations implementing more advanced simulation
techniques. To this aim, ATK provides two kinds of hook functions, which are
called either before or after the force calculation. The former type, called
\lstinline!pre_step_hook!, may modify e.g. positions, cell vectors, or velocities
of the current configuration, whereas the latter type, called
\lstinline!post_step_hook!, can be used to modify the forces and stress. Thus, a
\lstinline!pre_step_hook! can for instance be used to implement custom constraints,
apply strain to the simulation cell, or rigidly move atoms.
The \lstinline!post_step_hook! can be used to add external contributions or
to implement bias potentials to the regular interaction forces. For example
the system can be biased to explore configuration space more rapidly, as in
metadynamics. In ATK-2017, the metadynamics technique is included via a
\lstinline!post_step_hook! interface to the {\it PLUMED} package \cite{Tribello2014}.
Besides modifying the current state of the system, hook functions can be
employed to perform and store custom on-the-fly measurements at arbitrary
intervals. One advantage of this hook function framework is that the user can
directly implement such hook functions in the Python simulation script without
having to re-compile the code. Moreover, due to the standardized API there is
no the need for the user to learn specific implementation details of the MD
loop. ATK is shipped with some pre-defined hook functions, implementing e.g.
thermal transport via reverse non-equilibrium molecular dynamics (RNEMD)
\cite{MullerPlathe1997}.

It is often useful to define constraints on the atomic or other generalized
coordinates during an MD simulation. Two of the most commonly encountered
constraints are implemented in ATK: fixing the Cartesian coordinates of an
atom and allowing a group of atoms to only move rigidly. It is also possible
for the user to define their own custom constraints. Both the predefined and
custom constraints are evaluated in the same part of the MD loop (cf. Fig.
\ref{fig:mdloop}).

\section{Optimization}
\label{sec:optimization}

\subsection{Optimization Algorithms}

There are two optimization algorithms implemented in ATK: the fast inertial
relaxation engine (FIRE) \cite{Bitzek2006} and the limited-memory
Broyden-Fletcher-Goldfarb-Shanno (L-BFGS) quasi-Newton optimizer
\cite{Nocedal2006}.

The fast inertial relaxation engine (FIRE) uses a modified version of molecular
dynamics to locate a minimum energy configuration. The velocities at each step
are a weighted combination of the velocity at the previous step and the current
forces. The time step and the weighting factor, on the previous step's
velocities, increase as long as the velocities continue to point in a descent
direction (i.e. the angle between the velocity and the forces does not exceed
90 degrees). If the velocities are not a descent direction, then the timestep
and weights are reset to an initial safe value. For a complete description of
the algorithm see Ref.~\cite{Bitzek2006}.

L-BFGS is a quasi-Newton optimization algorithm, which unlike other
quasi-Newton methods, is able to treat problems of many variables because the
memory usage and computational cost at each step are controllable
\cite{Nocedal1980,Nocedal2006}. The implementation of L-BFGS in ATK makes use of
an automatic preconditioning scheme to determine the initial inverse Hessian at
each step. In the first step, the second derivative along the force is
estimated using finite differences. In all subsequent steps, it is chosen
according to Eqn.~2.4 in Ref.~\cite{Morales2000}. The preconditioning scheme
typically ensures that the proposed L-BFGS step reduces the function value
without the use of a line search. If the function value increases, then a
backtracking line search is used to ensure that progress is made.

L-BFGS is superior to FIRE for most optimization problems. In one benchmark
study on Lennard-Jones clusters, FIRE required 3.5 times more steps on average
than L-BFGS to reach convergence \cite{Chill2014Benchmark}. In another benchmark
comparing optimization algorithms for finding minimum energy pathways, L-BFGS
required 1.5--3.9 times fewer iterations than FIRE for NEB calculations on
surfaces \cite{Sheppard2008}.

\subsection{Geometry Optimization}

ATK features geometry optimization for isolated and periodic systems. For
molecules and clusters, geometry optimization involves minimizing the atomic
forces. While, for crystals, the shape and size of the unit cell as well as
any external pressure (possibly anisotropic) must also be taken into account.

The approach to optimizing crystals in ATK is based upon
Ref.~\cite{Sheppard2012}. The main idea is to express changes to the system as
a combined vector of atomic and strain coordinates. The strain expresses
changes to the lattice vectors relative to the initial configuration. By
combining the coordinates together into a single vector it is possible to
simultaneously optimize the atomic positions and the lattice vectors. The
optimization algorithm is given a combined force vector which will bring the
system into hydrostatic equilibrium,
\begin{equation} \label{eqn:bulk-forces} \mathbf{F}_\mathrm{simultaneous} =
  (-\nabla V(\mathbf{r}),
  -(\mathbf{\sigma}_\mathrm{cauchy}-\mathbf{\sigma}_\mathrm{external})
\frac{\Omega}{L\sqrt{n}}), \end{equation} where $-\nabla V(\mathbf{r})$ is the
usual atomic force vector, $\mathbf{\sigma}_\mathrm{cauchy}$ is the internal
stress tensor (calculated from the atomic interactions),
$\mathbf{\sigma}_\mathrm{external}$ is the external stress tensor, $\Omega$ is
the volume of the initial unit cell, $L$ is the average distance between atoms
in the initial unit cell, and $n$ is the total number of atoms. The factor of
$\Omega/L\sqrt{n}$ is included as a rescaling coefficient to ensure that the
stress, which has units of pressure, scales similarly to the atomic forces.
The rescaling coefficient is calculated during the first optimization step and
kept constant during subsequent steps.

\subsection{Reaction path optimization}

ATK has a state of the art NEB \cite{Henkelman2000} and climbing image
NEB \cite{Henkelman2000climbing} implementation that includes a newly developed
algorithm for generating initial guesses \cite{Smidstrup2014}, rapidly convergent L-BFGS
optimization, and per-image parallelization for scaling to large system sizes.

The initial set of replicas can be obtained in one of two ways: linear
interpolation between the end points or the image dependent pair potential
(IDPP) \cite{Smidstrup2014}. As The IDPP avoids unphysical starting guesses,
this method often leads to an initial band that is closer to the minimum energy
path (MEP) than a linear interpolation and it can typically reduce the required
number of optimization steps by a factor 2.

In some implementations the projected NEB forces for each image are optimized
independently, however, in that case the L-BFGS algorithm is known to
behave poorly \cite{Sheppard2008}.
In ATK, the NEB forces for each image are combined into a single vector:
$\mathbf{F}_\mathrm{NEB} \in \mathbb{R}^{3mn}$, where $m$ is the number of images
and $n$ is the number of atoms.
This combined approach is much more efficient when used with L-BFGS, and has been
referred to as the global L-BFGS (GL-BFGS) method \cite{Sheppard2008}.

\subsection{Adaptive Kinetic Monte Carlo}

Adaptive kinetic Monte Carlo (AKMC) is an algorithm for modeling the long
timescale kinetics of solid-state materials \cite{Xu2008,Chill2014AKMC}. The goal is to
construct a kinetic Monte Carlo (KMC) simulation by automatically locating the
relevant reaction mechanisms and calculating the reaction rates using harmonic
transition state theory (HTST) \cite{EON}.

For a given configuration AKMC involves 3 steps: (1) locate all kinetically
relevant product states; (2) determine the saddle point between the reactant
and product states; (3) select a reaction using Monte Carlo.

In ATK step 1 is performed using high temperature MD. At regular intervals the
MD simulation is stopped and a geometry optimization is performed. If the
geometry optimization converges to a new geometry then a reaction has occured.
This procedure is repeated until all relevant reactions are found (within an
user specified confidence). For details of the confidence calculation see
refs.~\cite{Chill2014AKMC,Aristoff2016}. The saddle point geometry for each
reaction is then determined by performing a NEB calculation for each reaction.
A reaction is then selected using MC and the system evolves to the
corresponding product configuration and the entire procedure repeated. Further
details of the AKMC implementation will be presented in a separate publication
\cite{Chill2017}.

\section{The ATKPython Platform}

In order to facilitate the integration of physical simulation methods in
Python, ATK provides \atkpython{}, an extension of the regular Python package.
Beyond the standard Python functionality, \atkpython{} automatically imports
the relevant modules which implement all ATK-functionality.
As mentioned in the previous section, the back-engines
of ATK-ForceField and other performance intensive parts, such as DFT
calculators, are efficiently implemented in C and C++. For a seamless
integration, \atkpython{} uses Python front-end classes, which automatically
call interface modules to these back-engines. Furthermore, \atkpython{} provides
a built-in \lstinline!PhysicalQuantity! module to conveniently handle and
convert physical quantities including their units. The implementation of
physical quantities is based on the Numpy-array structure, and \atkpython{}
provides support for almost all native Numpy-array operations with physical
quantities.
Configurations are intuitively set up by
specifying the coordinates and elements of the atoms, as well as details of
the simulation cell, for bulk, surface and device configurations.
When using bonded force fields, the bond connectivity of the configuration can
be explicitly specified or automatically detected based on interatomic distances
and covalent radii.

Calculators are defined by creating an instance of the desired calculator
class, e.g. a \lstinline!LCAOCalculator()! for DFT with linear combination of
atomic orbitals (LCAO). When constructing the objects, parameters can be given
to specify various calculator properties.
A simulation using ATK-ForceField can
be set up by specifying a pre-defined potential set object and initializing
the \lstinline!TremoloXCalculator! object with this potential set, e.g.
\begin{lstlisting}
potential_set = StillingerWeber_Si_1985()
calculator = TremoloXCalculator(parameters=potential_set)
\end{lstlisting}
which invokes the Stillinger-Weber potential for silicon \cite{Stillinger1985}.
If the desired force field is not listed among the pre-defined potentials,
one can set it up manually from the provided classes of interaction functions
in TremoloX.
To this aim, one first specifies an empty \lstinline!TremoloXPotentialSet! object,
which acts as container for the various components of the potential.
\begin{lstlisting}
potential_set = TremoloXPotentialSet('New potential')
\end{lstlisting}
Then, one needs to add the particle type objects, one for each element
contained in the system, e.g.
\begin{lstlisting}
potential_set.addParticleType(ParticleType(
  symbol='O',
  charge=-0.5*elementary_charge
))
\end{lstlisting}
One can choose to add different properties to the each type,
e.g. partial charges or the atomic Lennard-Jones coefficients.
To distinguish between different particle types of the same element, as required
e.g. in biomolecular force fields \cite{mackerell2004empirical},
particle types can be created such that they represent only a specified group of
atoms, marked by a given selection tag.
In the next step, one needs to set up objects for each potential function that
should be included in the potential.
The objects are typically initialized with the element or particle type
symbols between which the potential should act, the potential parameters themselves,
and some technical parameters, such as inner and outer cut-off radii.
An \atkpython{} object defining a Morse potential between silicon and oxygen,
for instance, could be set up as follows.
\begin{lstlisting}
morse_potential = MorsePotential(
    'Si',
    'O',
    r_0=2.1*Ang,
    k=2.0067*1/Ang,
    E_0=0.340554*eV,
    r_i=6.0*Ang,
    r_cut=7.5*Ang),
)
potential_set.addPotential(morse_potential)
\end{lstlisting}
Each of these potential function objects needs to be added to the
\lstinline!TremoloXPotentialSet! object.
If electrostatic interactions should be taken into account, a coulomb solver
can be added, which by default uses the partial charges specified with the
particle types, although atom-specific charges can be given as well.
To be used in a simulation, the calculator object must be attached to
the configuration object.

Various properties, such as the total energy, the forces, or the stress,
can be calculated via analysis objects, which return the corresponding
properties as a physical quantity in the appropriate units independent
of which type of calculator has been selected.
Beyond the basic properties, ATK also provides more complex analysis objects to
obtain, e.g. the phonon bandstructure, elastic constants, or local atomic
stress of a given configuration with attached calculator. Some of the analysis
objects, such as electron bandstructure or electron transmission spectrum,
depend on the electronic structure, meaning that they are only compatible with
quantum mechanical or semi-empirical calculators.

An MD simulation is set up by first defining the integration method, i.e. the ensemble and the algorithm
that is used to simulate that ensemble, as an object of
the desired integrator class. Here, depending on the chosen method, various
thermostat parameters, as well as the initial velocities, can be invoked when
setting up the integrator object.
\begin{lstlisting}
method = NVTNoseHoover(
    time_step=1*femtoSecond,
    reservoir_temperature=300*Kelvin,
    thermostat_timescale=100*femtoSecond,
    heating_rate=0*Kelvin/picoSecond,
    initial_velocity=MaxwellBoltzmannDistribution(300*Kelvin),
)
\end{lstlisting}
If no parameters are specified, default values are used, which generally work
well for a large range of simulation purposes.
Instead of a global temperature value, all NVT methods also accept a list
of \lstinline!(tag_name, temperature)!-tuples, in order to apply local thermostats
to the group of atoms marked by the corresponding tag.

Finally, the MD simulation over the desired number of integration steps is started
by calling the \lstinline!MolecularDynamics()! function.
Snapshots of the system are stored every \lstinline!log_interval! steps and written to
a trajectory file in {\it HDF5}, {\it NetCDF}, or {\it XYZ} format.
The entire MD block in the Python script may look like the following:
\begin{lstlisting}
md_trajectory = MolecularDynamics(
    configuration,
    constraints=[FixAtomConstraints([0, 1])],
    trajectory_filename='Silicon_NVT.hdf5',
    steps=5000,
    log_interval=100,
    pre_step_hook=None,
    post_step_hook=None,
    method=method
)
\end{lstlisting}

When all MD steps are done, the resulting \lstinline!md_trajectory! object
will contain all saved images and its associated properties, such as
potential energies. Thus, the simulation script can be arbitrarily extended,
by analyzing the \lstinline!md_trajectory! object, using built-in or custom MD
analysis objects. Note that in the HDF5-format the MD trajectory object is not stored in memory
but as a file on disk. When quantities are extracted from the final
trajectory, only the queried data is loaded into memory, which means that
even large trajectories, whose size exceeds the total amount of available
memory, can be analyzed.

To implement advanced functionality via hook functions as discussed in section~\ref{sec:dynamics},
one has to define the hook, either as a function, or as a callable Python class.
The parameters handed to such a function or method must be the step number, the current
simulation time, the configuration, and the atomic forces and the stress.
The configuration object, as well as the arrays containing forces and stress can
be modified inside the function and the simulation will continue with the new values.
Simple hook functions could look as in the following examples.
Here, the \lstinline!pre_step_hook! shifts the z-component of the position of the
first atom, while the \lstinline!post_step_hook! tethers the same position to a
fixed point using a harmonic bias potential.
\begin{lstlisting}
def pre_hook_example(step, time, conf, forces, stress):
    positions = conf.cartesianCoordinates()
    # Add an offset to the z-position of the first atom.
    positions[0, 2] += 0.1*Ang
    conf.setCartesianCoordinates(positions)

 def post_hook_example(step, time, conf, forces, stress):
    # Add a bias force to the force acting on the first atom
    # in the z-direction.
    positions = conf.cartesianCoordinates()
    bias_force = -0.5*eV/Ang**2*(positions[0, 2] - 2.0*Ang)
    forces[0, 2] += bias_force
\end{lstlisting}
Further examples for hook functions can be found in Sec.~\ref{sec:examples}.

\section{Workflow and Analysis}

The simulation workflow of typical MD simulation can be set up entirely in the
VNL graphical user interface. VNL is based on the \atkpython{} language, and
each tool in VNL can interpret and generate Python scripts, thus, it is
possible to seamlessly shift from the GUI to the scripting language. Most of
the main functionality in \atkpython{} is also available in VNL. VNL is
developed around a plugin concept which makes it easy to extend and add new
functionality. Plugins can be downloaded and installed from an add-on server
and the majority are available as source code, thus, they are easy to modify
or extend with new user defined functionality.

\begin{figure*}
\begin{center}
  \includegraphics[width=0.95\textwidth,clip]{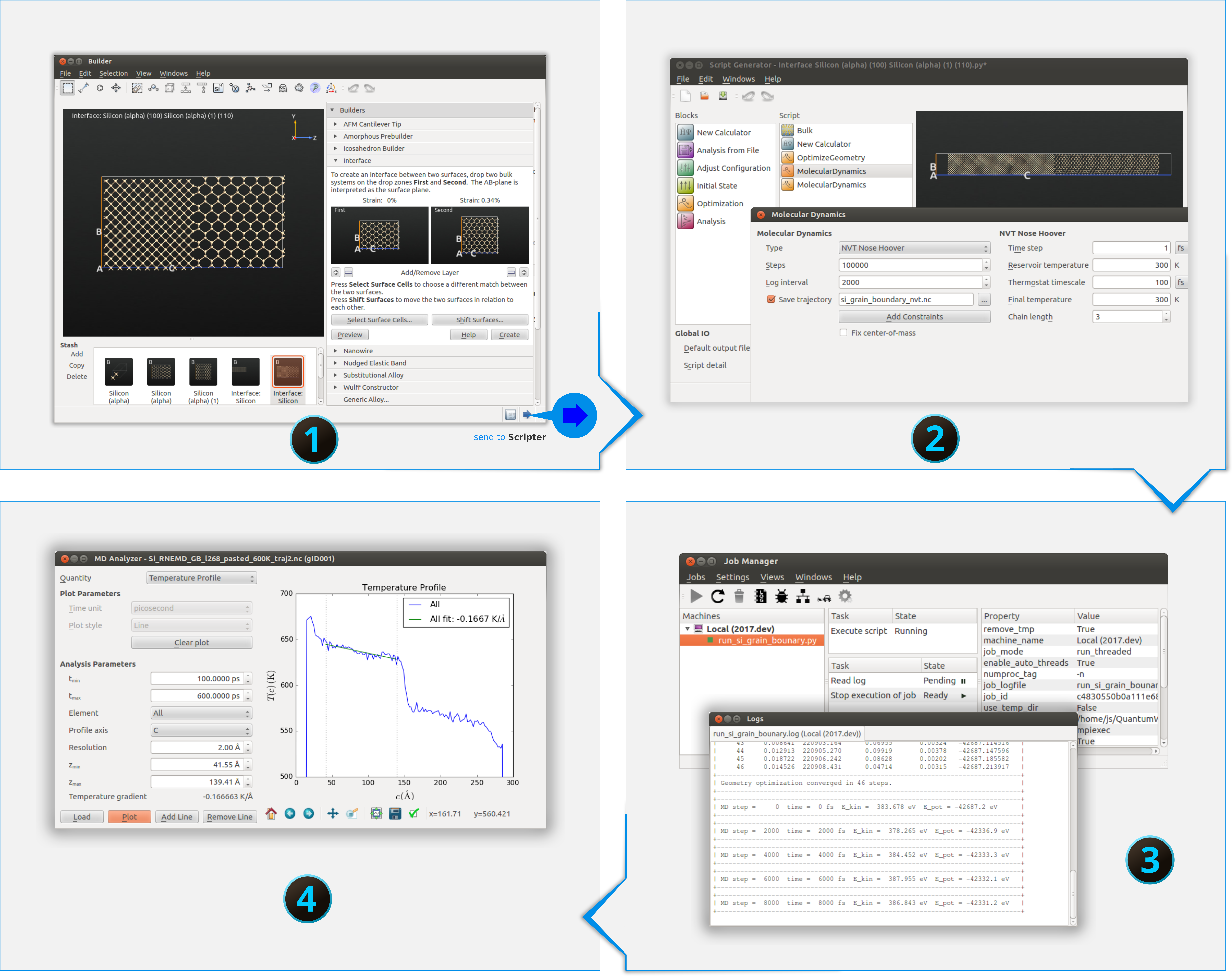}
  \caption{Example of a typical workflow for an MD simulation in VNL-ATK:
  The initial configuration is prepared using the plugins in the Builder,
  the simulation protocol is defined in the Script Generator,
  the simulation is started and monitored via the Job Manager,
  and the MD Analyzer is used to analyze the simulation results.}
  \label{fig:vnlworkflow}
\end{center}
\end{figure*}

A typical workflow to set up an MD simulation in VNL is presented in Fig.
\ref{fig:vnlworkflow} with screenshots of the corresponding tools. It starts
with setting up the atomic configuration in the {\em Builder} tool, using the
Builder databases, combined with Builder plugins to e.g. repeat, cleave, or
strain the structure, add, remove, or shift atoms, or build interfaces between
different structures. To set up the \atkpython{} simulation protocol, the
configuration is sent to the {\em Script Generator}, where a calculator, an
optimization or one or more MD simulations, as well as various analysis
objects can be added and their parameters set. The simulation script can
either be sent to the {\em Job Manager}, from which the calculation is run on
the local machine or on a remote cluster, or to the {\em Editor}. The Editor
allows for modifying or extending the Python script before the calculation is
run.

VNL provides various graphical analysis tools, which can
be used to visualize and analyze the simulation with respect to a broad range
of properties. At first, VNL provides the {\em Movie Tool}, which shows a movie
of the trajectory and simultaneously displays the corresponding graph of
potential, kinetic, and total energy, as well as the system temperature over
simulation time. High quality graphics and more advanced visual modifications,
such as changing the graphical properties of the atoms, e.g. by coloring the
atoms by forces or velocities, can be carried out in the {\em Viewer} tool.
Quantitative analysis of the simulation can be carried out using the various
analysis objects in VNL, based on the final structure after the MD or
optimization run. Here, the user can for instance calculate the local crystal
structure or the local stress of each atom, and visualize the results by
coloring the atoms in the Viewer. To extract averaged or time-dependent
properties from the MD trajectory itself, the {\em MD-Analyzer} can be used.
This tool provides a broad range of analysis quantities, such as radial
distribution function, neutron scattering function, angle and coordination
number distribution, vibrational density of state, mean-square displacement,
void-size distribution and many more.

\section{Simulation Examples}
\label{sec:examples}

In this section we provide examples of some MD simulations using ATK-ForceField.
All these examples can be run under the free academic license version of ATK.

\subsection{Interfacial thermal conductance}

\begin{figure}
\begin{center}
  \includegraphics[width=0.7\textwidth,clip]{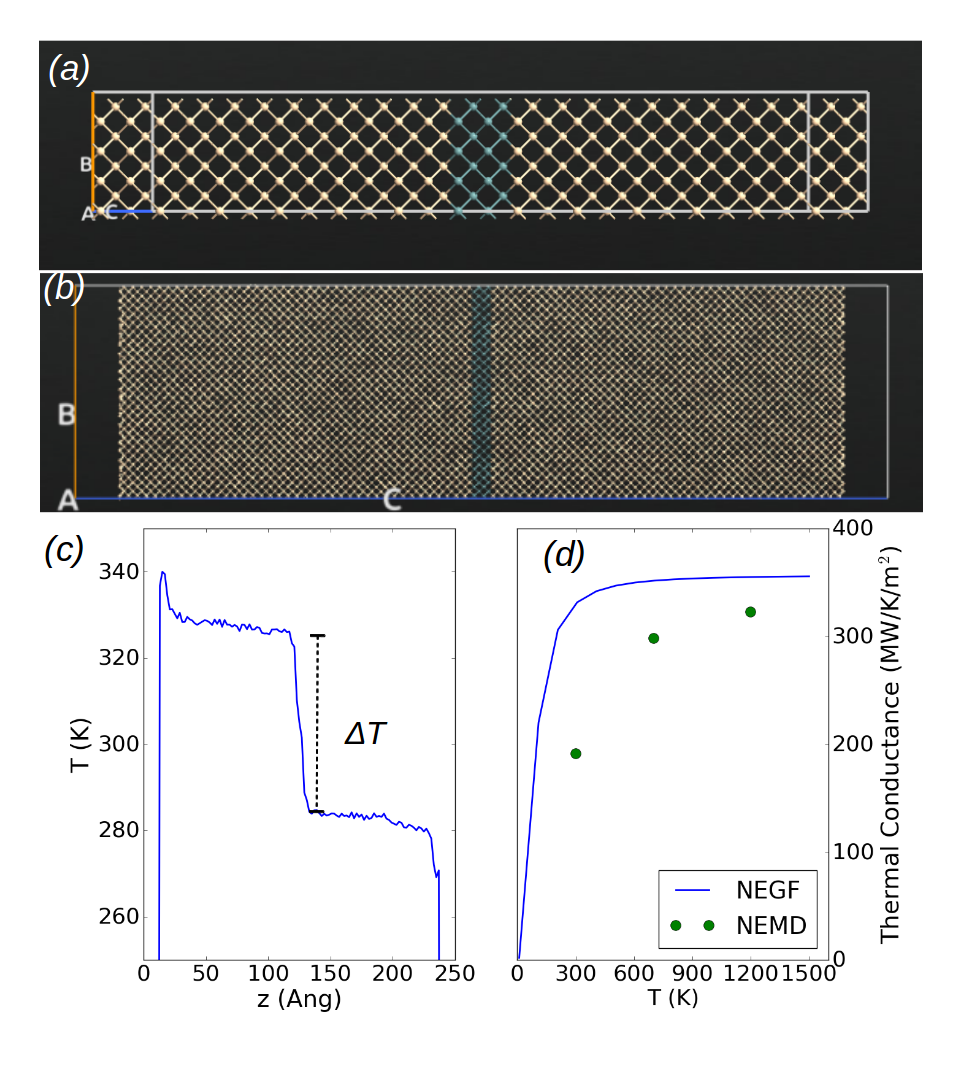}
  \caption{{\it (a)}: Device configuration used to simulate the thermal conductance
  through a germanium layer in a silicon crystal via NEGF,
  {\it (b)}: bulk configuration used in NEMD,
  {\it (c)}: the temperature profile at 300 Kelvin, with the temperature drop at the
  interface marked by the dashed line,
  {\it (d)}: interfacial thermal conductance at various temperature.
  The NEGF results are depicted by the solid line, and the NEMD results are shown as dots.}
  \label{fig:conductance}
\end{center}
\end{figure}

Thermal conductance through an interface can be simulated in different ways.
Two common approaches are the non-equilibrium Greens functions (NEGF) technique
to calculate the phonon transmission spectrum \cite{Markussen2009} and
non-equilibrium molecular dynamics (NEMD) based simulations in which a thermal
flux is imposed in the system \cite{MullerPlathe1997, Schelling2002}. Both
approaches have different advantages and shortcomings. NEGF based phonon
transmission only accounts for the elastic scattering of phonons and neglects
contributions from phonon-phonon scattering. While NEMD simulations include
both elastic and inelastic scattering, the technique is often limited by small
simulations cells, which not only neglects the effects of long-wavelength
phonons \cite{Howell2012} but also introduces temperature gradients that are
larger than the experimental conditions. Moreover, quantum effects, which
become important at low temperatures, are neglected. ATK provides a unique
possibility to use both techniques with the same calculators, and thus to
exactly compare the results.

In this example, we compare both techniques by
considering a simple model interface system: a silicon crystal with a single
layer of germanium atoms. The atomic interactions are modeled using a Tersoff
potential \cite{tersoff89}. For the NEGF-based phonon transmission, we use a
device configuration as displayed in Fig. \ref{fig:conductance} (a). The
phonon transmission is calculated as described in Ref.~\cite{Markussen2009} and
evaluated to obtain the thermal conductance as function of temperature. The
result is shown as the solid line in Fig.~\ref{fig:conductance} (d).

The NEMD simulations are performed on a bulk configuration containing a slab
geometry with free surfaces (cf. Fig. \ref{fig:conductance} (b)). The
terminating unit cell layers on each side are used the heat source and heat
sink, similar to the approach used in Ref.~\cite{Shen2014}.

Note the larger silicon crystal leads in the NEMD simulation compared to the
NEGF simulation. This is because in MD simulations the cell has to explicitly
accommodate all phonon modes which should be included in the thermal
conductance calculations, whereas in the NEGF approach the bulk character of
the semi-infinite leads is taken into account via the the q-points.

After equilibrating the system in the NVT ensemble, the NEMD simulations are
carried out in the NVE ensemble using a time step of 0.5~fs. To impose a
thermal flux, we employ the reverse NEMD algorithm
\cite{MullerPlathe1997, Nieto-Draghi2003}, which works via exchanging the
momenta of the hottest atom in the heat sink and the coldest atom in the heat
source. This NEMD functionality is readily available in ATK as a pre-defined
hook class, leaving only few parameters such as the exchange interval and the
tag-names of the heat source and sink regions to be specified.

The \atkpython{} script for an NEMD simulation is shown in the \ref{app:nemd}.

The temperature profile is calculated over a period of 300~ps, after discarding
the initial 200~ps and plotted using the MD-Analyzer tool in VNL (see
Fig. \ref{fig:conductance} (c)). At the location of the germanium layer a
pronounced temperature drop $\Delta T$ is visible. The interfacial thermal
conductance is obtained, based on Fourier's law, via
\begin{equation}
\kappa = \frac{\dot Q}{A \Delta T} \, ,
\end{equation}
where $\dot Q$ is the average thermal flux, which is printed at the end of the
simulation, and $A$ is the cross-sectional area of the interface.

The results for our model interface system are displayed in
Fig.~\ref{fig:conductance}. The thermal conductance obtained by NEMD is
slightly smaller than the results of the NEGF calculation. This can to a some
extent be attributed to finite size effects in the NEMD configuration, which
suppresses the contribution of some long-wavelength modes. In fact, some test
simulations with very long systems (up to 5 times longer leads on both sides)
showed that the NEMD-conductance increases to values slightly larger than the
NEGF results for all considered temperatures. However, for the scope of this
example, we did not perform an exhaustive convergence study with respect to
the system size. Importantly, we note that both methods yield values that are
consistent with each other.

\subsection{Vapor Deposition}

Vapor deposition is another area where MD simulations can be used to understand
the underlying microscopic processes and the effect of experimental parameters.

For technical reasons, most MD codes do not allow the number of atoms to change
during a single MD simulation. This makes depositions simulations challenging.
A user must run an initial MD simulation, then add new atoms or molecules in a
script outside the MD-code, prepare and write the updated input files, start a
new MD simulation, and repeat until the required timescale is reached. In ATK,
however, this can be achieved in a single Python script, by adding an outer
loop around the MD function, in which newly deposited atoms or molecules are
added to the current configuration at each cycle. Another advantage of this
approach, compared to a built-in deposition functionality which is offered e.g.
by the LAMMPS package, is that the user can use Python scripting
to exactly specify how the deposition is supposed to take place. In this
example, we demonstrate a vapor deposition simulation of selenium
molecules onto a crystalline substrate using a slightly modified Stillinger-Weber
potential \cite{oligschleger_1996}. To account for the polydispersity of
selenium vapor, the deposited molecules are drawn from a distribution of
ring-like selenium molecules of different sizes. In a recent study, we have
shown that such simulations can help elucidate the structural differences in
vapor-deposited films compared to melt-quenched amorphous selenium
\cite{Goldan2016}.

\begin{figure}
\begin{center}
  \includegraphics[width=0.5\textwidth,clip]{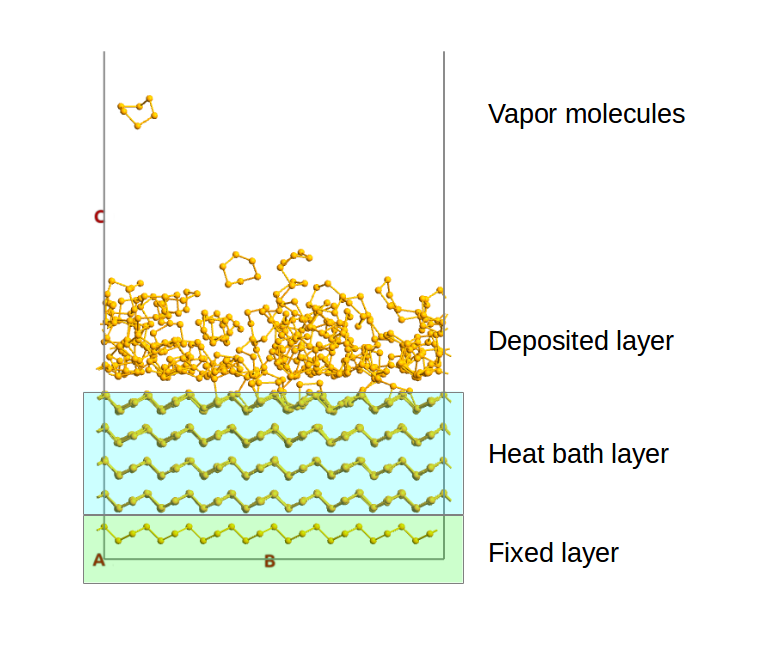}
  \caption{Snapshot from a deposition simulation.
  The different regions of the systems are annotated in the picture.}
  \label{fig:deposition1}
\end{center}
\end{figure}

A schematic picture of the simulation setup is shown in
Fig.~\ref{fig:deposition1}. The starting configuration consists of a slab model
of the (100) surface of a trigonal selenium crystal. In this substrate
configuration, two regions are selected and marked via tags: The bottom layer,
which will be fixed in the entire deposition simulation to mimic an infinitely
extended perfect crystal substrate, and the substrate region above this fixed
layer. In the MD simulation, a thermostat will be applied only to that region
to remove excess thermal energy from the substrate, while leaving the ballistic
dynamics of the incoming molecules unaltered.

Schematically this looks as
shown in the code snippet in the \ref{app:deposition}.
First, a function (only the signature is shown) is defined, that draws a
new molecule from a given distribution of molecules. Then the deposition loop
is started. In each loop cycle, a new molecule is drawn and its center of mass
is shifted to a random lateral position at a fixed height above the substrate.
Thermal velocities, corresponding to the desired vapor temperature, are
assigned making sure that the center-of-mass velocity is directed towards the
surface. The new molecule is inserted into the last MD configuration and a new
MD simulation over 20 ps is started from the updated configuration.
\begin{figure}
\begin{center}
  \includegraphics[width=0.95\textwidth,clip]{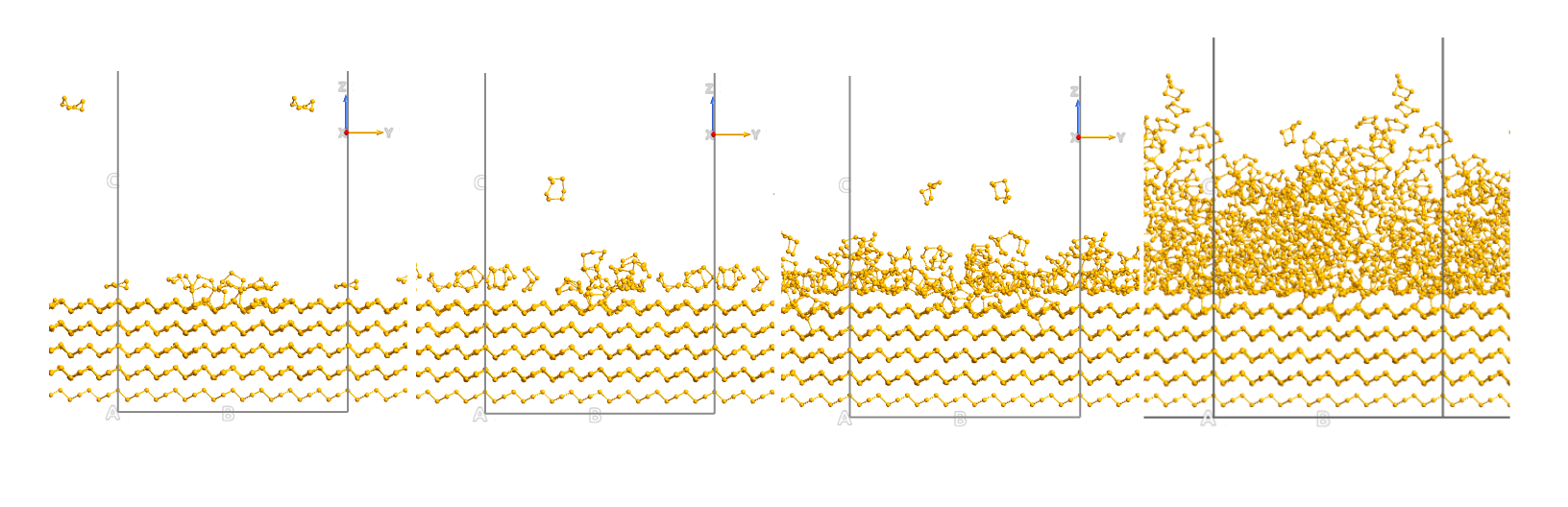}
  \caption{Snapshots from a different stages of the selenium vapor deposition simulation
  ({\it left to right}):
  8 molecules, 20 molecules, 60 molecules, and 200 molecules.
  The simulation cell is indicated by the black lines.}
  \label{fig:deposition2}
\end{center}
\end{figure}

The simulation is run for 200 deposition cycles producing a 25~\AA\ thick
amorphous film. Snapshots of the intermediate stages of the deposition
simulation are shown in Fig.~\ref{fig:deposition2}. Upon adsorption at the
substrate surface, a fraction of the initially ring-like molecules begin
polymerizing to form longer chain-like structures. In contrast to experiments,
where a precise determination of the exact molecular topology is nearly
impossible, the simulation results provide atomistic insight into the initial
structure immediately after deposition. In this case, it is possible to
determine the ratio between atoms in ring- or chain-like topologies. The
simulation results suggest that the vapor-deposited film exhibits
predominantly ring-like structures \cite{Goldan2016} and that the ratio is
temperature dependent. This is in contrast to the results obtained from the
simulation of melt-quenched amorphous selenium where the dominating molecular
structure is chain-like.

\subsection{Creep simulation of a copper polycrystal}

\begin{figure*}
\begin{center}
  \includegraphics[width=0.8\textwidth,clip]{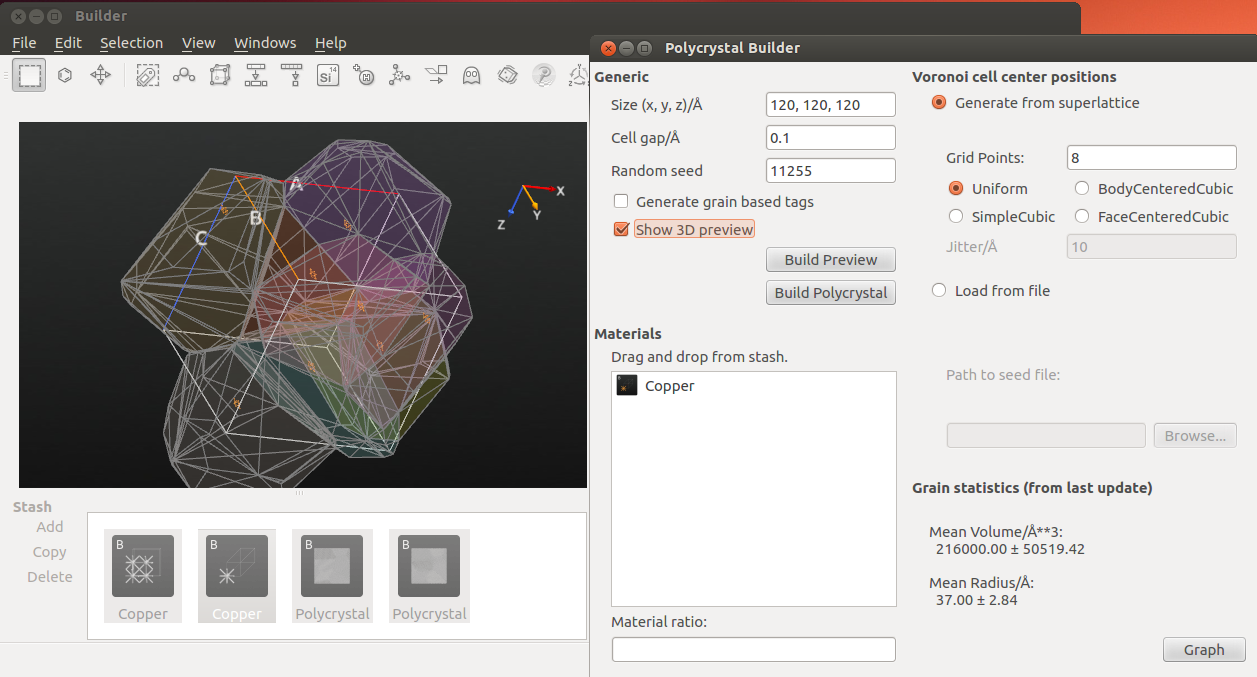}
  \caption{Screenshot of the polycrystal builder in the VNL-Builder tool.
  The 3D-view shows the grain seeds and the grain boundaries as colored polyhedra.}
  \label{fig:polycrystal}
\end{center}
\end{figure*}

Even in microscopic devices, many components are not composed of perfect single crystals,
but instead contain small crystalline grains, which form a polycrystalline material.
Thus, the properties of such a material are not only dominated by the properties of
the corresponding crystal itself but by an interplay with the grain boundaries.

\begin{figure}
\begin{center}
  \includegraphics[width=0.95\textwidth,clip]{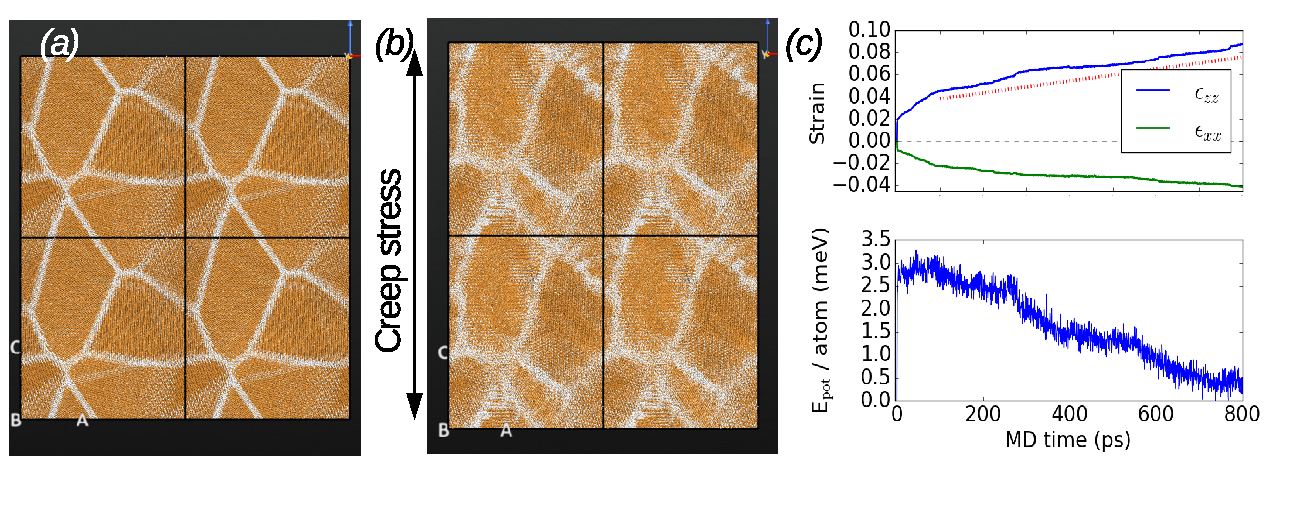}
  \caption{Grain structure of the polycrystal before {\it (a)} and after {\it (b)} the creep simulation.
  The atoms in a local FCC arrangement are colored in yellow, whereas the atoms in grain boundaries are marked white.
  For better visualization, the picture shows a 2x2 repetition, the actual simulation cell is indicated by the black lines.
  {\it (c):} Strain relaxation in tensile ($z$) and lateral ($x$) direction,
  and potential energy relaxation during the creep simulation as a function of simulation time.
  The slope of $\epsilon_{zz}(t)$, representing the creep rate $\dot{\epsilon}$
  is shown by the dashed line.}
  \label{fig:creep}
\end{center}
\end{figure}

Here, we present the simulation of a creep experiment of a copper polycrystal
as an example of how such systems can be simulated in VNL and ATK. The initial
configuration is set up in the polycrystal builder within VNL (see Fig.
\ref{fig:polycrystal})\footnote{The Polycrystal builder plugin is part of the
SCAITools package which can be installed in VNL using the AddOn-manager.
Besides generating the grain centers using random numbers, ATK has an {\tt
autoseed} tool, which can generate grain seeds according to a given grain size
distribution, using a genetic algorithm
\cite{Barker.Bollerhey.Hamaekers:2015}.
The generated seed positions can be loaded into the polycrystal builder and the
polycrystal builder fills the grains with the given crystal materials.}.
It is possible to specify the size, the
estimated number of grains, and the algorithm by which the grain seeds are
determined. In this example, we use a cell volume of $12 \times 12 \times 12 \,
{\rm nm}^3$. Eight initial seed positions are generated based on a uniform random
distribution, and the grains are filled by copper FCC crystals with random
orientations. The resulting grains have an average volume of 216~nm$^3$
corresponding to an average diameter of 7.4~nm.

To describe the interactions in the system we use an EAM potential \cite{Mishin2001}.
Figure~\ref{fig:creep} (a) and (b) show the grain structure of the initial polycrystal
and after the creep simulation. The grain boundaries are visualized via the
{\it Local Structure} analysis in VNL, which assigns each atom a local crystal
structure by analyzing the arrangement of the neighboring atoms
\cite{Stukowski2012}. Before the actual creep simulation, the configuration is
equilibrated at a temperature of 500K and zero pressure. For the creep
simulation, a tensile stress of 1.5 GPa is applied in the z-direction, while
the lateral directions remain coupled to zero pressure. The creep simulation is
run for 800 ps using a time step of 2 fs.
The code snippet in the \ref{app:polycrystal} shows the \atkpython{}
script for this simulation block.
Since the configuration contains around 150 000 atoms, saving snapshots of the
entire system at a high frequency will produce huge trajectory files. On the
other hand, it may be necessary to analyze some observables, such as the
strain, at a much higher time resolution. As shown in the example script, this
can be achieved via a hook function, which measures and stores the desired
quantities, in this case the strain, every 100 MD steps, while regular
snapshots of the system are stored only every 25 000 steps.

The strain relaxation as function of the simulation time is shown in
Fig.~\ref{fig:creep} (c). Almost immediately after applying the tensile
stress, the system responds by elongating by about 2\%. This deformation is
primarily elastic. In the following simulation time, the strain increases
more slowly, which is characteristic of creep behavior. Fig.~\ref{fig:creep}
(c) shows the corresponding changes in potential energy per atom. The initial,
mainly elastic deformation is accompanied by an increase in energy, while the
creep relaxation results in a continuous decrease in potential energy.
However, due to the small number of grains, the strain does not exhibit a
nearly linear increase, as typically expected for polycrystalline
materials, but a more step-like behavior. The average creep rate
$\dot\varepsilon=5\cdot10^7$ s$^{-1}$ is depicted by the dotted line in Fig.
\ref{fig:creep} (c).

\section{Outlook}

In this paper, we have presented the dynamics and optimization functionality
in ATK, including the ATK-ForceField module for empirical potentials. One of
the main strengths of the ATK suite is its full integration with the Python
scripting language which makes it very flexible and allows for defining
complex workflows without the need for recompiling any code. Due to the
implementation of the performance-intensive parts in C/C++, this flexibility
is obtained without sacrificing computational efficiency and the performance
of ATK-ForceField is on par, or in some cases faster, than other state-of-the-art
classical MD codes. The ATK is seamlessly integrated with the graphical
user interface VNL, which makes it easy to set up and interpret \atkpython{}
scripts. In this respect, the usability of the VNL-ATK suite is unparalleled
and it can both lower the barrier for beginners in MD simulations as well as
make it easier for experts to develop and implement new MD functionality.
A recent example, highlighting the advantage of integrating calculators with
different levels of accuracy, shows how ATK can be used to
combine MD simulations using ATK-ForceField with advanced electronic transport
calculations to efficiently obtain the electronic mobility and conductivity of
various bulk systems and nano-wires \cite{Markussen2017}.

ATK and VNL are under active development by QuantumWise. ATK-ForceField is
developed in collaboration with Fraunhofer SCAI. A number of features not
available in ATK-2017, in particular supporting TremoloX's native hybrid
MPI/OpenMP-parallelization in ATK, are planned for ATK-2018.
Furthermore, a toolbox for fitting potential parameters to ab-initio calculations
is under development. We particularly expect such workflows for fitting
parameters to benefit from synergies present in the
\atkpython{} framework, e.g. calculators with increasing levels of
accuracy are readily available to provide reference observables, and
a broad range of optimization libraries are natively included in Python modules,
such as scipy.


\section{Acknowledgements}
This work was funded in parts by the German Federal Ministry for Education and
Research and the Innovation Fund Denmark under the Eurostars projects E!6935
ATOMMODEL and E!9389 MULTIMODEL.

\appendix

\section{Code Examples}

Here, we provide some snippets containing the \atkpython{} code for the simulation
examples in section~\ref{sec:examples}.

\subsection{NEMD Simulations of Thermal Transport}
\label{app:nemd}
The following snippet shows how to run a non-equilibrium MD simulation to
obtain the thermal conductance through an impurity layer of germanium in a silicon crystal.
The exchange of momenta according to the RNEMD method is invoked as a {\em hook function} object
in the MD simulation, which is supplied as a pre-defined class in \atkpython{}.
It is initialized by passing the two tag names referring to the groups of atoms in the
heat source and sink, as well as the interval at which momenta are exchanged.

\begin{lstlisting}[label=lstconductance]
# Add tags to atoms in the heat source and heat sink
bulk_configuration.addTags('heat_source', heat_source_indices)
bulk_configuration.addTags('heat_sink', heat_sink_indices)

# Define the hook function which invokes the exchange of momenta.
momentum_exchange_hook = NonEquilibriumMomentumExchange(
    configuration=bulk_configuration,
    exchange_interval=200,
    heat_source='heat_source',
    heat_sink='heat_sink'
)

# Set up the MD integrator method.
method = NVEVelocityVerlet(
    time_step=0.5*femtoSecond,
    initial_velocity=ConfigurationVelocities()
)

# Run the RNEMD simulation for 1 000 000 steps.
md_trajectory = MolecularDynamics(
    bulk_configuration,
    constraints=[],
    trajectory_filename='Si_w_Ge_NEMD.hdf5',
    steps=1000000,
    log_interval=1000,
    post_step_hook=momentum_exchange_hook,
    method=method,
)

# Print the average thermal flux in the NEMD simulation
nlprint(momentum_exchange_hook)
\end{lstlisting}

\subsection{Vapor Deposition Simulation of Selenium}
\label{app:deposition}
The following snippet shows how a vapor-deposition simulation of selenium
molecules onto a crystalline substrate can be set up in \atkpython{}.
The function to select a new molecule from a given distribution is only
shown schematically and can be implemented as desired for the given application.
The central part of the script is a loop over all deposition events.
In each iteration a new molecule with thermal velocities directed towards the
surface is added to the configuration. This is followed by an MD
simulation in which the deposited molecule can adsorb and the configuration can
equilibrate for a short time.

\begin{lstlisting}
# Define helper function to randomly draw a molecule from
# a molecular distribution with given probabilities.
def selectFromDistribution():
    ...
    return coordinates

# Loop over the deposition events.
for i_deposit in range(deposition_cycles):
    # Generate the coordinates for a new vapor molecule.
    new_positions = selectFromDistribution()
    mol_size = len(new_positions)

    # Shift the molecule to a random lateral position at fixed height.
    new_positions[:,0] += numpy.random.uniform()*Lx
    new_positions[:,1] += numpy.random.uniform()*Ly
    new_positions[:,2] += deposition_height

    # Set random velocities at vapor temperature.
    thermal_vel = (T_vapor*boltzmann_constant/Selenium.atomicMass())**0.5
    new_velocities = numpy.random.normal(size=(mol_size, 3))*thermal_vel
    com_velocity = new_velocities[:,2].sum()/mol_size

    # The z-component of the velocity should point towards the substrate.
    if com_velocity > 0.0*Ang/fs:
      new_velocities[:, 2] -= 2.0*com_velocity
    new_elements = [Selenium]*mol_size

    # The new molecule is added to the current configuration.
    bulk_configuration._insertAtoms(
        elements=new_elements,
        positions=new_positions,
        velocities=new_velocities,
    )

    # Run an MD simulation depositing the atom towards the surface.
    md_trajectory = MolecularDynamics(...)

    # Extract the final snapshot as starting configuration
    # for the next deposition cycle.
    bulk_configuration = md_trajectory.lastImage()
\end{lstlisting}

\subsection{Creep Simulation of a Copper Polycrystal}
\label{app:polycrystal}
The following snippet shows how creep simulation of a copper polycrystal can be
run using \atkpython{}. Here, a hook function is used to calculate and store the current
strain in the system.

\begin{lstlisting}
# Define a hook class to store the time-dependent strain.
class StrainHook(object):
    def __init__(self, interval):
        self._interval = interval
        self._cells = []

    def __call__(self, step, time, configuration, forces, stress):
        """ Apply the hook in the MD simulation. """
        if (step % self._interval == 0):
            cell = configuration.primitiveVectors().inUnitsOf(Ang)
            self._cells.append(cell)

    def strain(self):
        """ Return the strain in z-direction. """
        cells = numpy.array(self._cells)
        strain_zz = cells[:, 2, 2]/cells[0, 2, 2] - 1.0
        return strain_zz

# Set up the barostat with anisotropic coupling to tensile stress of 1.5 GPa.
# The lateral target stress is set to zero.
method = NPTMartynaTobiasKlein(
    time_step=2*femtoSecond,
    reservoir_temperature=500*Kelvin,
    reservoir_pressure=[0, 0, -1.5]*GPa,
    initial_velocity=ConfigurationVelocities(),
)

# Calculate and store the strain every 100 steps.
strain_hook = StrainHook(interval=100)

# Run the creep simulation.
md_trajectory = MolecularDynamics(
    bulk_configuration,
    constraints=[],
    trajectory_filename='Cu_polycrystal_creep.hdf5',
    steps=400000,
    log_interval=25000,
    pre_step_hook=strain_hook,
    method=method
)
bulk_configuration = md_trajectory.lastImage()

# Obtain the time-dependent strain stored in the hook.
strain_zz = strain_hook.strain()
\end{lstlisting}

\section*{References}

\providecommand{\newblock}{}

\end{document}